# Absolute frequency of $^{87}$Sr at 1.8×10$^{−16}$ uncertainty by reference to remote Primary Frequency Standards

**Nils Nemitz, Tadahiro Gotoh, Fumimaru Nakagawa, Hiroyuki Ito, Yuko Hanado, Tetsuya Ido, and Hidekazu Hachisu**

National Institute of Information and Communications Technology,
 4-2-1 Nukui-kitamachi, Koganei, Tokyo, 184-8795, Japan

E-mail: Nils.Nemitz@nict.go.jp

**Abstract**

The optical lattice clock NICT-Sr1 regularly reports calibration measurements of the international timescale TAI. By comparing measurement results to the reports of eight Primary Frequency Standards, we find the absolute frequency of the $^{87}$Sr clock transition to be $f(\text{Sr}) = 429\,228\,004\,229\,873.082(76)$ , with a fractional uncertainty of less than $1.8 \times 10^{-16}$ approaching the systematic limits of the best realization of SI second. Our result is consistent with other recent measurements and further supported by the loop closure over the absolute frequencies of $^{87}$Sr, $^{171}$Yb and direct optical measurements of their ratio.

## 1. Introduction

Optical clocks, frequency standards that probe atomic references in the optical frequency regime, now reach uncertainties close to 1 part in $10^{18}$, enabling new science such as relativistic geodesy, dark matter searches and investigations of changes in fundamental constants [1-6]. Yet, all of our daily interaction with International Atomic Time TAI and its derived timescales rely on continuity above all else. To ensure this, the Consultative Committee for Time and Frequency (CCTF), instituted by the Comité International des Poids et Mesures (CIPM), has defined a set of milestones [7] to be reached before the definition of the SI second will change from the currently referenced microwave transition in caesium ($^{133}$Cs) to an optical reference with significantly greater accuracy.

One of these milestones is contribution of optical clocks to the steering of TAI according to the procedures established for $^{87}$Rb microwave standards [8]. Following pioneering work at LNE-SYRTE [9], our own NICT-Sr1 and several other optical lattice clocks [10,11], have been officially recognized as Secondary Frequency Standards (SFS). After recognition in December 2018, NICT-Sr1 has contributed to the steering of TAI by reporting measurement results for use in the monthly Circular T issued by the International Bureau of Weights and Measures (BIPM). As a clock utilizing the $^1S_0 \to {}^3P_0$ transition in $^{87}$Sr, the weight of these contributions is now limited by an uncertainty of $4 \times 10^{-16}$ assigned to neutral strontium as a Secondary Representation of the Second (SRS) [7] based on previous measurements (References [9,12-15] among others) of the absolute frequency.

Here we present a new evaluation of the Sr absolute frequency, referencing the ensemble of caesium-based Primary Frequency Standards (PFS) that submit TAI evaluations to BIPM. The results are consistent with the latest published results [16] for a comparison to local standards. Our evaluation not only tests the performance of the clocks, but in employing the same infrastructure used in the steering of TAI, it directly supports a greatly reduced SRS uncertainty of $1.8 \times 10^{-16}$ or below.

## 2. Methods

The full measurement chain is described in this section. Figure 1 gives a schematic overview of tracing the frequency of NICT-Sr1 to the international timescales for comparison to remote PFSs.

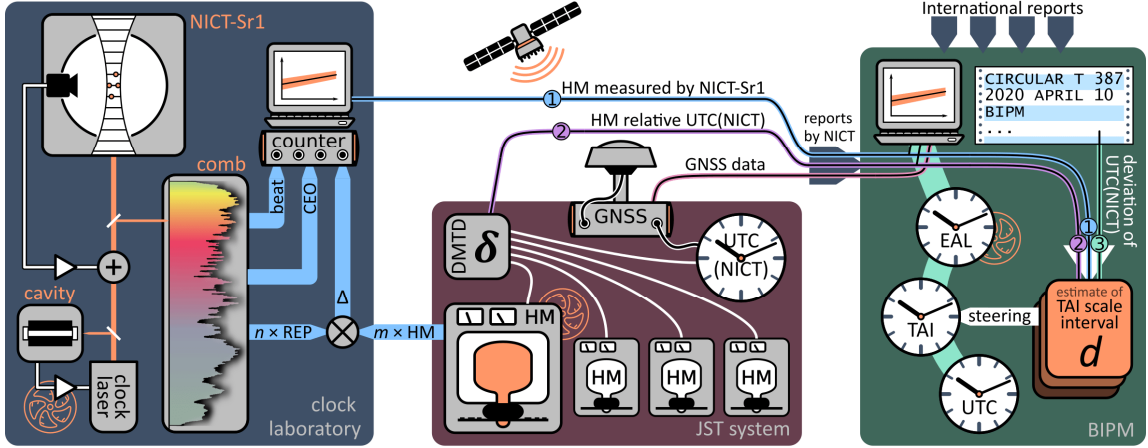

**Figure 1.** Remote comparison of NICT-Sr1 to international timescales for absolute frequency evaluation. ① The optical lattice clock NICT-Sr1 characterizes the frequency of a hydrogen maser (HM). ② This HM is also continuously compared to NICT's local timescale UTC(NICT) within the Japan Standard Time (JST) system. ③ The Bureau of Weights and Measures (BIPM) evaluates submitted data of global navigation satellite systems (GNSS) to find the time deviation of UTC(NICT) from the calculated UTC scale. The evolution of this deviation provides the frequency difference. Combining ①-③ characterizes the frequency of UTC (and the identical frequency of TAI) through measurements of NICT-Sr1, typically expressed as the deviation of the TAI scale interval $d$. Calculated for each frequency standard reporting to BIPM, the collective estimates of the deviation are used to steer TAI. We use the same values to compare the frequency NICT-Sr1 to remote frequency standards operated by international institutes. See text for further details.

## 2.1 Optical clock

The optical lattice clock NICT-Sr1 typically traps 1 000 $^{87}$Sr atoms in a vertical, 1-dimensional optical lattice created by the standing wave of a laser tuned to a magic wavelength near 813 nm, where it provides tight confinement with negligible disturbance to the atomic resonance frequency [17]. By detecting the excitation state of the trapped atoms, a clock laser is then stabilized to the transition frequency of approximately 429 228 004 229 873 Hz. Measurements destroy the atomic sample, limiting the clock to periodic evaluations with a cycle time of typically 1.5 s. A reference cavity made of ultra-low-expansion glass thus serves as a flywheel oscillator. With atomic measurements providing corrections for changes in the cavity's resonant frequency and drift rate, the clock laser provides a continuous optical reference with a frequency-noise limited fractional instability that falls with averaging time $\tau$ as $\sigma_y(\tau) = 7 \times 10^{-15} \, (\tau/\text{s})^{-1/2}$. The systematic uncertainty evaluation of NICT-Sr1 is described in reference [15]. Table 1 shows the nominal uncertainty budget with an overall fractional uncertainty $u_b = 5.4 \times 10^{-17}$, largely limited by the evaluation of residual lattice light shifts. Each measurement submitted to BIPM is accompanied by a report that includes a specific uncertainty budget [18].

**Table 1.** Nominal uncertainty budget of NICT-Sr1.

| Effect | Correction ($10^{-17}$) | Uncertainty ($10^{-17}$) |
|---|---|---|
| Blackbody radiation | 513.9 | 3.0 |
| Light shifts | 2.6 | 3.8 |
| DC Stark shift | 0.1 | 0.5 |
| Quadratic Zeeman shift | 52.0 | 0.3 |
| Hot and cold collisions | 2.2 | 2.3 |
| Servo error | 0.0 | 0.6 |
| Total | 570.8 | 5.4 |
| Gravitational redshift | –834.1 | 2.2 |
| Total (with redshift) | –263.3 | 5.9 |

The servo error uncertainty represents the possibility of a persistent deviation of the clock laser frequency from the atomic resonance, as may occur due to incomplete cancellation of the cavity drift. The mean deviation observed in data accumulated over four years is $3.7(4.6) \times 10^{-18}$. We thus reassign a conservative uncertainty of $\sqrt{0.37^2 + 0.46^2} \times 10^{-17} = 0.59 \times 10^{-17}$. This represents errors that affect all measurements equally, while random instabilities are handled as part of the statistical contribution.

*2.2 Intermittent clock operation*

International timescale comparisons rely on satellite links, either by Two-Way Satellite Time-and-Frequency Transfer (TWSTFT) or by Precise Point Positioning (PPP) over global navigation satellite systems (GNSS), as used at NICT. The accuracy of these links relies on long uninterrupted measurements, modelled by an uncertainty falling with $\tau^{-0.9}$ [19] to reach $2 \times 10^{-16}$ after approximately 30 days in the case of NICT. Current optical clocks are not yet reliable enough for uninterrupted operation of this length. We thus use a hydrogen maser (HM) that is part of the Japan Standard Time (JST) system [20] to serve as additional flywheel oscillator during gaps in the measurements.

We characterize HM behaviour based on the phase differences between multiple HMs (Anritsu Corporation), continuously monitored at 1 s intervals by a multi-channel dual-mixer time difference (DMTD) system [21]. As discussed in section 2.7, the long-term contribution is best described by an Hadamard deviation of $\sigma_H^2(\tau) = a_0^2 + a_2^2(\tau/1\ \mathrm{d})^2$, where $a_0 = 2.1 \times 10^{-16}$ represents the floor of flicker frequency noise (FFN), and the contribution from $a_2 = 1.7 \times 10^{-17}$ is often referred to as flicker-walk frequency modulation (FWFM) [22]. Over several years of available data, we find the corresponding $\sigma_H(\tau) \propto \tau$ behaviour to be a better descriptor than the more commonly used frequency random-walk, which appears as $\sigma_H(\tau) \propto \tau^{1/2}$.

Uncertainties introduced by unobserved stochastic HM behaviour are thus much smaller than the uncertainties of the GNSS link not only for short interruptions in clock operation, but also for intervals of multiple days. This opens the path to increased accuracy by simultaneous comparison to the multiple PFSs that calibrate the international timescales [23,24], at a manageable level of experimental effort: We deliberately operate NICT-Sr1 in a pattern of intermittent measurements [25,26] and use the flywheel HM to extend the effective period of evaluation. Similar approaches have also been adopted in other optical clock measurements that rely on remote links [10,11,16,27].

*2.3 Reference maser evaluation*

During the operation of NICT-Sr1, an optical frequency comb compares the optical reference to the signal of the reference HM at $f_{HM} = 100$ MHz. All measurements are performed phase-coherently without deadtime to provide efficient suppression of phase noise in the HM signal. For averaging times $\tau > 10$ s, the observed instability approximately follows $\sigma_y(\tau) = 4 \times 10^{-14}\ (\tau/\mathrm{s})^{-1/2}$ (figure 3) until the drift of the HM frequency limits the stability. The drift rate is very consistent [28], and the instability of drift-removed data matches expectations from the flicker-floor coefficient $a_0 = 2.1 \times 10^{-16}$.

To minimize measurement noise and avoid counting errors, the frequency comb measurements are performed as illustrated in figure 1. The HM signal is multiplied by a factor $m$ using a phase-locked dielectric resonator oscillator (PLDRO, Nexyn Corporation NXPLOS). The $n$-th harmonic of the comb repetition rate $f_r$ is detected by a high-speed photodetector (Discovery Semiconductors DSC40S) at the output of the same amplified and frequency-broadened optical path that is used to generate the beat signal with the clock laser. By downmixing this signal with the output of the PLDRO, the counted frequency (indicated by Δ in figure 1) becomes

$$f_c = n\,f_r - m\,f_{HM} = 50\ \mathrm{MHz} \tag{1}$$

for a typical configuration with $m = 92$, $n = 37$ and $f_r = 250$ MHz. This relaxes the requirements on the counter accuracy by a factor of $(n\,f_r)/f_c = 185$. Detector and PLDRO operate in a temperature-stable environment and protected from air currents. A zero-deadtime multichannel counter (K&K Messtechnik FXE) measures $f_c$ simultaneously with the clock laser's beat $f_b$ with the nearest comb line $n_{clk}$ and the comb's carrier-envelope offset $f_{CEO}$, obtained from an $f - 2f$ interferometer. The clock laser frequency, as it appears at the comb, can then be determined as

$$f'_{clk} = f_{CEO} + n_{clk} f_r + f_b \quad . \tag{2}$$

The measurements of $f_{CEO}$ and $f_b$ are also used to detect disturbances of the clock laser frequency and cycle-slips of the comb lock, so that affected data can be removed from the evaluation. When the comb is locked to the clock laser, the small in-loop error of $f_b$ allows direct detection of even single-cycle slips over the 1 s measurement interval. When the comb is locked to the HM for greater robustness, we operate an additional tracking oscillator that is phase-locked to the $f_b$ signal with narrow control bandwidth. Counted as $f_t$ on a separate counter channel, laser disturbances will then cause different-valued miscounts and yield $|f_t - f_b| \geq 1$ Hz. This allows unreliable data to be rejected despite a noise band of $f_b$ that is wider than 1 Hz.

The primary concern in terms of systematic measurement errors are then phase shifts induced by thermal effects on the cables delivering the HM signal from the adjacent building that is home to the JST system. Presuming a daily temperature cycle, this may lead to persistent errors if measurements typically occur at the same time of day. We test for these effects in the data set of a nearly continuous ten-day measurement by binning the data by time-of-day and examine a sliding six-hour window to obtain maximum sensitivity to diurnal effects. We observe no excursions beyond statistical expectations. For the specific time window of 14:00 to 20:00, where most measurements are performed, we set a limit of $\hat{u}_d = 7.95 \times 10^{-17}$ by adding the observed deviation from the mean in quadrature with the statistical uncertainty and then take $u_{b/lab} = \hat{u}_d$ as the systematic uncertainty for the frequency link to the reference HM. For longer measurement intervals, cyclical phase shifts will have less effect, and we model this as a systematic uncertainty falling as $u_{b/lab} = \hat{u}_d/(T_i/6 \text{ h})$ for measurements with individual operating times $T_i > 6$ h.

Measurements before MJD 58400 acquired a lower (1 GHz) harmonic of the repetition rate directly from the comb oscillator, which was then frequency divided to 100 MHz for phase comparison to the HM reference. These measurements showed sensitivity to the thermal environment of the frequency divider. Additionally, the phase comparator was found to introduce a persistent error depending on the measured frequency difference $\Delta y \approx 10^{-12}$ as $\delta_{pc} = 4.6 \times 10^{-5} \Delta y$. This results from numerical errors in the original software version and a frequency error of the integrated timebase. Measurements relying on phase comparator data accommodate this by a larger systematic uncertainty $u_{b/lab} = 1 \times 10^{-16}$. As summarized in table 3, we now apply a retroactive correction for $\delta_{pc}$ based on our present understanding, but maintain the originally reported $u_{b/lab}$. After an initial thermalization for 12 h, a test at $\Delta y \approx 0$ finds counter and phase comparator measurements to agree to $\delta y = 6.6 \times 10^{-19}$ over 48 h.

*2.4 HM drift correction*

We consider the observed HM fractional frequency deviation $y_{HM}^{obs}(t)$ from the nominal frequency to consist of a constant linear drift $y_{HM}^{lin}(t)$ with superimposed noise of zero mean value. We seek the mean value $\bar{y}_{HM}$ over the chosen evaluation period $T$. As illustrated in figure 2, we approach the drift by fitting the observed subset of data as

$$y_{HM}^{lin}(t) = \bar{y}_{HM}^{obs} + a\,(t - \bar{t}) \quad , \tag{3}$$

where the observed mean value $\bar{y}_{HM}^{obs}$ corresponds to the value at the barycentre $\bar{t}$. Performing the fit relative to this point, which represents the mean of all observation times, provides the clearest separation of the uncertainties of the observed mean value and of the drift rate.

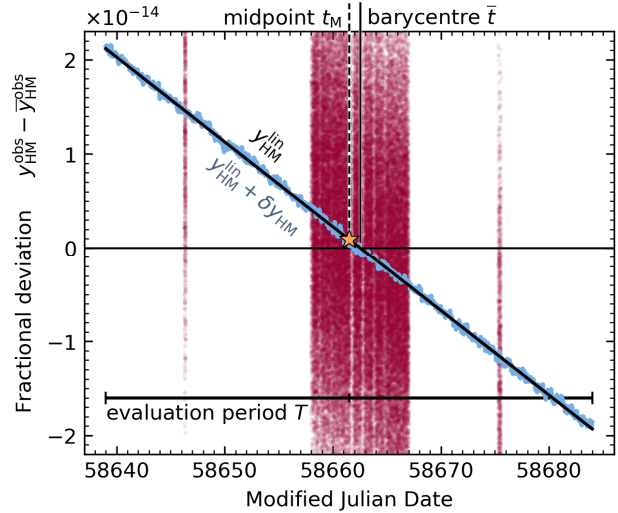

**Figure 2.** Determination of mean frequency $\bar{y}_{HM}$ from intermittent data. Observations $y_{HM}^{obs}$ binned at 10 s intervals (translucent points) are interpreted as a linear drift $y_{HM}^{lin}$ (black line) overlaid with stochastic noise. The observed mean value $\bar{y}_{HM}^{obs}$ corresponds to $y_{HM}^{lin}(\bar{t})$ at the barycentre $\bar{t}$. The mean value $\bar{y}_{HM}$ (star) over the evaluation period $T$ is found as $y_{HM}^{lin}(t_M)$ at midpoint $t_M$, with a correction given by the slope of the fit. In handling the deadtime between observations, slowly varying, non-WFN contributions are of particular concern. Such contributions are partially revealed in the hourly estimate $\delta y_{HM}$ obtained by comparison to the ensemble of here $N_{ens} = 3$ HMs. To obtain the best agreement of $y_{HM}^{lin} + \delta y_{HM}$ (blue-shaded line) with observations, $y_{HM}^{lin}$ is (re-)determined by a linear fit to $y_{HM}^{obs} - \delta y_{HM}$.

Analysis (see section 2.5) finds the instability at the original 1 s measurement interval to be dominated by phase noise. We bin the frequency data over 10 s intervals where this is suppressed to below the level of white frequency noise (WFN). The fit to the binned data can then be interpreted in the context of normally distributed, uncorrelated noise. Finding $\bar{y}_{HM}$ from $\bar{y}_{HM}^{obs}$ now requires a correction from the barycentre $\bar{t}$ to the midpoint $t_M$ of the evaluation period

$$\delta y_{mid} = a(t_M - \bar{t}) \approx \bar{y}_{HM} - \bar{y}_{HM}^{obs} \qquad (4)$$

that carries an uncertainty $u_{mid} = \sigma_a |t_M - \bar{t}|$, where the slope $a$ and its uncertainty $\sigma_a$ are obtained from the fit. The statistical uncertainty $u_{stat}$ of $\bar{y}_{HM}^{obs}$ is evaluated separately as below.

*2.5 Short-term measurement instability*

After subtracting the fitted linear drift, the Allan deviation of the residuals provides information on the instabilities of the HM and the measurement of its frequency. At short averaging times, where phase noise of the HM signal strongly contributes to the instability, the reference cavity serves as a stable flywheel oscillator, and in conjunction with the zero-deadtime frequency measurement avoids the aliasing effects [29] that typically accompany cyclic clock interrogation. The instability initially falls as approximately $\sigma_y(\tau) \propto \tau^{-1}$, before the slope follows $\sigma_y(\tau) \propto \tau^{-1/2}$ over averaging times $10\text{ s} < \tau < 10\,000\text{ s}$, consistent with WFN as the primary noise process. We extrapolate this trend to the total measurement time $T_t$ for the evaluation period. Figure 3 illustrates the observed instability, where we find $u_{stat} = \sigma_y(T_t) = 3.9 \times 10^{-17}$ for $T_t = 779\,853$ s of exemplary data acquired over ten days of near-continuous clock operation.

*2.6 Long-term HM instability*

The excess instability observed at long averaging time represents stochastic HM behaviour that we seek to quantify. It does not limit the accuracy of the measured mean frequency. However, similar behaviour during unobserved intervals leads to a deadtime uncertainty. As discussed in reference [14], extrapolating the frequency of a flywheel oscillator –characterized by a known noise power spectral density (PSD) $S_y$– from one distribution of measurements to another incurs an uncertainty of

$$u_{\text{stoc}}^2 = \int_0^\infty S_y(f) \, |G_\Delta(f)|^2 \, df \quad . \tag{5}$$

Here $G_\Delta(f)$ is the Fourier transform of the differential weighting function $g_\Delta(t) = g_1(t) - g_2(t)$, where normalized distributions $g_1(t)$ and $g_2(t)$ describe the two distributions, in this case representing observation intervals and the full evaluation period.

**Table 2.** Allan and Hadamard variances for power spectral densities $S_y(f)$ of different noise types. Allan and Hadamard variances were calculated based on references [22,30]. The latter also lists equivalent Allan variances. The coefficient $f_H$ represents a high frequency cutoff according to the bandwidth of the measurement, typically taken as $f_H = 1/(2\,\tau_0)$ for a measurement interval $\tau_0$. The Euler-Mascheroni constant is $\gamma \approx 0.577$.

| Noise Type | Abbrev. | $S_y(f)$ | Allan variance ($\sigma_y^2$) | Hadamard variance ($\sigma_H^2$) |
|---|---|---|---|---|
| White phase noise | WPN | $h_2 f^2$ | $\frac{3 f_H}{4\pi^2} h_2 \tau^{-2} = \sigma_y^2(\tau)$ | $\frac{5 f_H}{6\pi^2} h_2 \tau^{-2} = \sigma_H^2(\tau)$ |
| Flicker phase noise | FPN | $h_1 f$ | $\frac{3\gamma + \ln 4 + 3\ln(\pi f_H \tau)}{4\pi^2} h_1 \tau^{-2} = \sigma_y^2(\tau)$ | $\frac{10\gamma + \ln 48 + 10\ln(\pi f_H \tau)}{12\pi^2} h_1 \tau^{-2} = \sigma_H^2(\tau)$ |
| White frequency noise | WFN | $h_0$ | $\frac{1}{2} h_0 \tau^{-1} = \sigma_y^2(\tau)$ | $\frac{1}{2} h_0 \tau^{-1} = \sigma_H^2(\tau)$ |
| Flicker frequency noise | FFN | $h_{-1} f^{-1}$ | $2\ln(2)\, h_{-1} = \sigma_y^2(\tau)$ | $\frac{1}{2}\ln\left(\frac{256}{27}\right) h_{-1} = \sigma_H^2(\tau)$ |
| Random-walk freq. mod. | RWFM | $h_{-2} f^{-2}$ | $\frac{2}{3}\pi^2 h_{-2} \tau = \sigma_y^2(\tau)$ | $\frac{1}{3}\pi^2 h_{-2} \tau = \sigma_H^2(\tau)$ |
| Flicker-walk freq. mod. | FWFM | $h_{-3} f^{-3}$ | does not converge | $\frac{16}{6}\pi^2 \ln\left(\frac{3}{4}\cdot 3^{11/16}\right) h_{-3} \tau^2 = \sigma_H^2(\tau)$ |

We handle phase and white frequency noise as part of $u_{\text{stat}}$ to better account for excess noise in the measurement system or varying day-to-day performance. Long-term HM behaviour under the operating conditions at NICT is described by a model Hadamard variance

$$\sigma_H^2(\tau) = (2.1 \times 10^{-16})^2 + (1.7 \times 10^{-17}\,\tau/1\,\text{d})^2 \tag{6}$$

that accounts for FFN and FWFM. A corresponding PSD can then be calculated [30] from the sensitivity function of the Hadamard deviation [22] as

$$S_y^{\text{HM}}(f) = \underbrace{4.0 \times 10^{-32}\,(f/\text{Hz})^{-1}}_{\text{FFN}} + \underbrace{3.0 \times 10^{-45}\,(f/\text{Hz})^{-3}}_{\text{FWFM}} \quad . \tag{7}$$

The required relations are given in Table 2. In the presence of the steep slope of $S_y^{\text{HM}}$ near $f = 0$ Hz, zero-padding the differential weight $g_\Delta(t)$ improves the accuracy of the calculations when using a discrete Fourier transform to obtain $G_\Delta(f)$. We find the results in good agreement with a Monte Carlo analysis using random noise specified by $S_y(f)$ according to reference [31].

Early measurements used a less flexible stochastic model [15,26]. Although this provides sufficient accuracy for the limited and homogeneously distributed deadtime in these measurements, we recalculate $u_{\text{stoc}}$ according to the updated model (see Table 3).

## 2.7 HM ensemble evaluation

The HMs of the JST system are distributed across multiple rooms and operate in a well-controlled environment. We expect their stochastic behaviour to be independent. Long-term observation by DMTD comparisons confirm similar behaviour and instabilities. Starting from hourly data spanning multiple years, we calculate Hadamard variances for the frequency difference $\delta y^{a,b} = y_a - y_b$ of each pair of HMs $a$ and $b$. The longest common operating interval is used for this, and only frequency excursions $> 5 \times 10^{-14}$ are rejected as outliers. We average the variances weighted by the length of available data and divide by 2 to determine an ensemble-average instability for a single HM. We fit this according to equation (6) with an additional WFN term and find $\sigma_H^2(\tau) = \left(4.2 \times 10^{-14}/\sqrt{\tau/\text{s}}\right)^2 + (2.1 \times 10^{-16})^2 + (1.7 \times 10^{-17}\,\tau/1\,\text{d})^2$. The WFN contribution is in close agreement with the observations in measurements by NICT-Sr1.

Although flicker noise processes exhibit complex temporal correlations, they can be constructed from normally distributed contributions with arbitrary accuracy [31,32], such that we expect a reduced uncertainty $u_{\text{stoc}}/\sqrt{N_{\text{ens}}}$ for an ensemble of $N_{\text{ens}}$ equally weighted HMs. To make use of the ensemble stability, we look at the frequency difference $y^{\Delta}_{\text{HM\_1}}$ of the reference HM (here HM_1) from the ensemble mean of the HM frequencies $\hat{y}_{\text{HM}\_i}$ recorded by the DMTD system relative to UTC(NICT):

$$y^{\Delta}_{\text{HM\_1}}(t) = \hat{y}_{\text{HM\_1}}(t) - \frac{1}{N_{\text{ens}}}\sum_i \hat{y}_{\text{HM}\_i}(t) \tag{8}$$

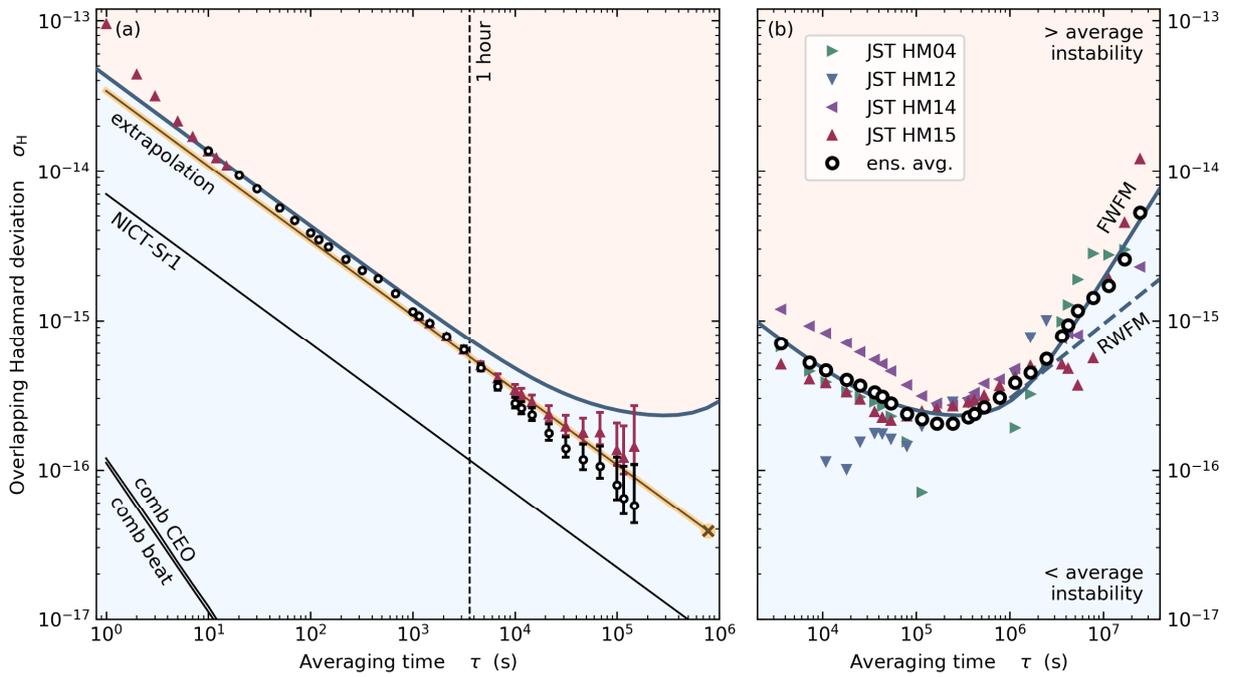

**Figure 3.** HM instability observed against NICT-Sr1 and within the JST HM ensemble. All results are shown as fractional overlapping Hadamard deviations to allow direct comparison. **(a)** Instability of JST HM15 in a near-continuous 10 d measurement. To account for drift of the HM, a linear fit was subtracted from the data before closing the gaps in clock operation. Open circles represent the overall instability after applying a correction for the non-linear behaviour of HM15 according to the estimate $\delta y_{\text{HM}}$. The (brown) extrapolation line illustrates $1/\sqrt{\tau}$ behaviour corresponding to WFN of $3.9 \times 10^{-17}$ (indicated by x) over all $T_t = 779\,853$ s of total data (90% of the 10 d interval). Red triangles show the uncompensated instability of HM15, which is consistent with the HM stability model shown by the curved blue line. Evaluated data is averaged over 10 s intervals, but a subset (shown for $\tau \leq 20$ s) was analyzed at 1 s timestep to illustrate the phase-noise contribution. NICT-Sr1 itself contributes an insignificant instability of $7 \times 10^{-15}/\sqrt{\tau/\text{s}}$. The measured instabilities of the comb CEO and beat with the clock laser are negligible. The vertical dashed line shows the 1 h averaging time of the JST data used to calculate ensemble correction and stability model. The instability reduction from the correction calculated from JST HM03, HM04, HM12, HM14 and HM15, is clearly visible beyond this point, and consistent with the expectation of $\sqrt{1/N_{\text{ens}}} \approx 0.45$. **(b)** HM stability model. The (blue) solid line (identical to (a)) shows the single-HM stability model fitted to the ensemble-average instability indicated by open circles. Triangles illustrate individual instabilities of JST HM04, HM12, HM14 and HM15, estimated by a four-corner-hat method [22]. Negative variance estimates result in missing points here, but the different HM characteristics are nevertheless visible. For $\tau > 10^6$ s, the observed behaviour does not match the common assumption of random-walk frequency modulation (RWFM, as indicated by dashed line), corresponding to $\sigma_H^2(\tau) \propto \tau$. A model following $\sigma_H^2(\tau) \propto \tau^2$, associated with flicker-walk frequency modulation (FWFM), provides better agreement.

We perform a linear fit $\tilde{y}^{\text{ens}}_{\text{HM\_1}}(t)$ over the evaluation period and subtract it from $y^{\Delta}_{\text{HM\_1}}(t)$ to eliminate both the mean value and the combined drift contributions. This defines

$$\delta y_{\text{HM}}(t) = y^{\Delta}_{\text{HM\_1}}(t) - \tilde{y}^{\text{ens}}_{\text{HM\_1}}(t) \tag{9}$$

as an estimate of the stochastic deviations of HM_1 from an ideal linear drift. This is subtracted from the observations to obtain ensemble-corrected data with reduced noise

$$y^{\text{NR}}_{\text{HM}} = y^{\text{obs}}_{\text{HM}}(t) - \delta y_{\text{HM}}(t) \ . \tag{10}$$

We then redetermine $y^{\text{lin}}_{\text{HM}}(t)$ of equation (3) from $y^{\text{NR}}_{\text{HM}}$, as in figure 2. Figure 3 confirms the resulting reduction in instability. As constructed, $\delta y_{\text{HM}}(t)$ integrates to zero over the full evaluation period, so that equation (10) only affects the results due to measurement deadtime. The resulting change in estimated $\bar{y}_{\text{HM}}$ is generally within the range given by $u_{\text{stoc}}$, and typically below $10^{-16}$ in magnitude. During the determination of $\delta y_{\text{HM}}(t)$ we calculate frequency differences and Hadamard deviations to exclude HMs with degraded stability from the ensemble calculations. Table 3 lists $N_{\text{ens}}$ for each evaluation.

## 2.8 Lab-side frequency determination

The uncertainty of the mean HM frequency $\bar{y}_{\text{HM}}$ over the chosen evaluation period is represented within the convention of the Circular T by the clock's statistical uncertainty $u_a$, its systematic uncertainty $u_b$ and the uncertainty $u_{\text{l/lab}}$ in the link between the frequency standard and the reference clock implemented by the HM. This consists of the contributions discussed in the previous sections:

$$u^2_{\text{l/lab}} = \underbrace{u^2_{\text{mid}} + u^2_{\text{stat}} + u^2_{\text{stoc}}/N_{\text{ens}}}_{u^2_{\text{a/lab}}} + u^2_{\text{b/lab}} \tag{11}$$

Although the Circular T has only recently begun reporting separate values for $u_{\text{a/lab}}$ and $u_{\text{b/lab}}$, the reports of NICT-Sr1 have always included a listing of $u_{\text{b/lab}}$ as discussed in section 2.3, which we treat as correlated across repeated measurements. The contributions grouped as $u^2_{\text{a/lab}}$ are expected to average according to the standard error of the mean.

## 2.9 Frequency link to TAI

The phase of the reference HM is continuously compared to UTC(NICT), which is evaluated against UTC (and thus TAI, which is identical in frequency) through a GNSS link [33] with a timing instability of now typically $\sigma^{\text{link}}_a = 0.3$ ns [18]. Time differences are calculated by BIPM for UTC 0:00 of modified Julian dates (MJD) ending in the digit 4 or 9. Relative frequencies can then be calculated over the intervening periods with a conventional uncertainty [19] of

$$u_{\text{l/Tai}} = \frac{\sqrt{2}\,\sigma^{\text{link}}_a}{T_0} \Big/ \left(\frac{T}{T_0}\right)^{0.9} \tag{12}$$

where $T$ is the length of the evaluated period and $T_0 = 5 \times 86\,400$ s represents the 5 d interval. The exponent of 0.9 empirically accounts for long-term instabilities that exceed the fundamental $1/T$ behavior expected for the white phase noise described by $\sigma^{\text{link}}_a$. Evaluated over typical 25 d to 35 d intervals, frequency standards commonly report $u_{\text{l/Tai}} \approx 2 \times 10^{-16}$. The uncertainty of the HM−UTC(NICT) comparison is negligible.

## 2.10 Frequency trace to primary standards

For NICT-Sr1, the Circular T data completes the frequency chain

$$\frac{f(\text{TAI})}{f(\text{SI})} = \frac{f(\text{TAI})}{f(\text{HM})} \cdot \frac{f(\text{HM})}{f(\text{Sr})} \cdot \frac{f(\text{Sr})}{f(\text{SI})} \ . \tag{13}$$

The results are expressed in terms of the fractional deviation $d$ of the scale interval $1/f(\text{TAI})$ from the SI second

$$d = \frac{f(\text{SI})}{f(\text{TAI})} - 1 \approx -\left(\frac{f(\text{TAI})}{f(\text{SI})} - 1\right) = -y_{\text{TAI}} \quad . \tag{14}$$

Since $d$ is typically of magnitude $1 \times 10^{-15}$, the approximation yields no loss in accuracy. Equivalent values are reported for all frequency standards contributing to TAI calibration, and we can trace from $f(\text{Sr})$ as measured by NICT-Sr1 to the results of a PFS by equating the results for $d$ to obtain an expression for the Sr clock frequency in terms of the SI second:

$$\frac{f(\text{Sr})}{f(\text{SI})} = \frac{f(\text{Sr})}{f(\text{HM}_\alpha)} \cdot \frac{f(\text{HM}_\alpha)}{f(\text{TAI})} \cdot \frac{f(\text{TAI})}{f(\text{HM}_\beta)} \cdot \frac{f(\text{HM}_\beta)}{f(\text{PFS})} \cdot \frac{f(\text{PFS})}{f(\text{SI})} \tag{15}$$

The HMs used as flywheel oscillators for the Sr and PFS measurements are differentiated by α and β. Each ratio term introduces uncertainty as discussed in the following. Most also represent measurements over specific intervals $T$ that require extrapolation:

$$\frac{f(\text{Sr})}{f(\text{SI})} = \frac{f(\text{Sr})}{f(\text{HM}_\alpha,T_{\alpha1})} \cdot \frac{f(\text{HM}_\alpha,T_{\alpha1})}{f(\text{HM}_\alpha,T_{\alpha2})} \cdot \frac{f(\text{HM}_\alpha,T_{\alpha2})}{f(\text{TAI},T_{\alpha2})} \cdot \frac{f(\text{TAI},T_{\alpha2})}{f(\text{EAL},T_{\alpha2})} \cdot \frac{f(\text{EAL},T_{\alpha2})}{f(\text{EAL},T_{\beta2})}$$

$$\cdot \frac{f(\text{EAL},T_{\beta2})}{f(\text{TAI},T_{\beta2})} \cdot \frac{f(\text{TAI},T_{\beta2})}{f(\text{HM}_\beta,T_{\beta2})} \cdot \frac{f(\text{HM}_\beta,T_{\beta2})}{f(\text{HM}_\beta,T_{\beta1})} \cdot \frac{f(\text{HM}_\beta,T_{\beta1})}{f(\text{PFS})} \cdot \frac{f(\text{PFS})}{f(\text{SI})} \tag{16}$$

Here we extrapolate TAI by first converting to the underlying Échelle Atomique Libre (EAL) which evolves continuously (see figure 4), while TAI is steered to match the SI second scale interval by stepwise corrections.

We can rewrite the equation as a sum of small fractional deviations $y$ from the respective nominal values. For the Sr absolute frequency, the chosen nominal value is the 2017 CIPM recommendation for neutral $^{87}$Sr as SRS $f_{\text{SRS}}(\text{Sr}) = 429\,228\,004\,229\,873.0$ Hz [7], such that

$$f(\text{Sr}) = \left[y\left(\frac{\text{Sr}}{\text{SI}}\right) + 1\right] \cdot f_{\text{SRS}}(\text{Sr}) \quad . \tag{17}$$

Then

$$y\left(\frac{\text{Sr}}{\text{SI}}\right) = -\underbrace{\left(y\left(\frac{\text{HM}_\alpha}{\text{Sr}},T_{\alpha1}\right) + y\left(\frac{\text{HM}_\alpha,T_{\alpha2}}{\text{HM}_\alpha,T_{\alpha1}}\right) + y\left(\frac{\text{TAI}}{\text{HM}_\alpha},T_{\alpha2}\right)\right)}_{\text{reported as } d \text{ in Circular T}} - \underbrace{y\left(\frac{\text{EAL}}{\text{TAI}},T_{\alpha2}\right) + y\left(\frac{\text{EAL},T_{\alpha2}}{\text{EAL},T_{\beta2}}\right) + y\left(\frac{\text{EAL}}{\text{TAI}},T_{\beta2}\right)}_{\text{reported in feal-ftai and fpsfs-ftai}}$$

$$+ \underbrace{\left(y\left(\frac{\text{HM}_\beta}{\text{PFS}},T_{\beta1}\right) + y\left(\frac{\text{HM}_\beta,T_{\beta2}}{\text{HM}_\beta,T_{\beta1}}\right) + y\left(\frac{\text{TAI}}{\text{HM}_\beta},T_{\beta2}\right)\right)}_{\text{reported as } d \text{ in Circular T}} + \underbrace{y\left(\frac{\text{PFS}}{\text{SI}}\right)}_{0} \tag{18}$$

All this data is publicly available from the BIPM FTP server [18]. The abbreviations "feal−ftai'" and "fpsfs−ftai" refer to the reports "Difference between the normalized frequencies of EAL and TAI" and "Difference between PSFS frequency and TAI frequency".

For each evaluation of NICT-Sr1 as listed in table 3, we determine a value $y(\text{Sr/SI})$ for every PFS that reported a calibrating measurement with overlapping period. Seven of the evaluations were performed with frequent clock operation homogenously distributed over the full Circular T reporting period. These achieve low uncertainties from the link to TAI as well as from extrapolation to the PFS evaluation through EAL. Here we use the reported value of $d$ after applying a correction for the phase comparator error discussed in section 2.3. For five other evaluations, we make use of the improved HM characterization and extend the evaluation periods to more closely match the PFS data by recalculating $d$ from the reported time difference UTC−UTC(NICT) and our records of HM−UTC(NICT). The same calculations were used to add an additional period representing four unreported clock operations between MJD 58419 and 58449.

**Table 3.** Estimates of the TAI scale interval $d$ by NICT-Sr1. Period of Estimation is given as MJD. Results and uncertainty contributions are given in units of $10^{-15}$. The number of HM used in the ensemble correction $N_\text{ens}$ and the operating duty fraction of NICT-Sr1 over the total period is also given. Comparison with coincident PFS reports yield a fractional deviation of the Sr absolute frequency $y(\text{Sr/SI})$ and uncertainty $u(\text{Sr/SI})$ for each period, with weights and covariances for individual PFS contributions handled as described in the text. Notes: *a*: value of $d$ corrected for phase comparator error, *b*: previously unreported data, *c*: extended period of estimation

| Period of Estimation | $d$ ($\times 10^{-15}$) | $u_\text{a}$ ($\times 10^{-15}$) | $u_\text{b}$ ($\times 10^{-15}$) | $u_\text{a/lab}$ ($\times 10^{-15}$) | $u_\text{b/lab}$ ($\times 10^{-15}$) | $u_\text{l/Tai}$ ($\times 10^{-15}$) | $u$ ($\times 10^{-15}$) | $N_\text{ens}$ | duty fraction | $y(\text{Sr/SI})$ ($\times 10^{-15}$) | $u(\text{Sr/SI})$ ($\times 10^{-15}$) |
|---|---|---|---|---|---|---|---|---|---|---|---|
| 57474–57504 | −0.382[a] | 0.03 | 0.081 | 0.338 | 0.10 | 0.20 | 0.415 | 2 | 2.1% | +0.510 | 0.509 |
| 57504–57539 | −0.413[a] | 0.03 | 0.074 | 0.289 | 0.10 | 0.17 | 0.359 | 2 | 2.0% | +0.586 | 0.530 |
| 57539–57569 | −0.175[a] | 0.03 | 0.075 | 0.273 | 0.10 | 0.20 | 0.362 | 2 | 2.1% | +0.279 | 0.419 |
| 57569–57599 | −0.616[a] | 0.03 | 0.063 | 0.265 | 0.10 | 0.20 | 0.354 | 2 | 2.1% | −0.224 | 0.436 |
| 57599–57629 | −0.718[a] | 0.03 | 0.059 | 0.292 | 0.10 | 0.20 | 0.373 | 2 | 2.6% | +0.072 | 0.422 |
| 57629–57659 | −0.970[a] | 0.03 | 0.058 | 0.314 | 0.10 | 0.20 | 0.391 | 2 | 2.2% | −0.225 | 0.535 |
| 58149–58174 | −0.025 | 0.029 | 0.072 | 0.249 | 0.10 | 0.308 | 0.416 | 2 | 2.7% | +0.070 | 0.491 |
| 58419–58449[b] | +0.486 | 0.034 | 0.084 | 0.252 | 0.080 | 0.261 | 0.382 | 5 | 1.6% | −0.172 | 0.444 |
| 58449–58474[c] | +1.099 | 0.008 | 0.078 | 0.115 | 0.001 | 0.308 | 0.338 | 5 | 36.6% | +0.429 | 0.416 |
| 58479–58514[c] | +0.866 | 0.037 | 0.077 | 0.277 | 0.080 | 0.201 | 0.361 | 5 | 1.2% | +0.242 | 0.403 |
| 58514–58539[c] | +1.022 | 0.020 | 0.075 | 0.181 | 0.080 | 0.231 | 0.313 | 4 | 5.8% | +0.547 | 0.380 |
| 58639–58684[c] | +0.714 | 0.008 | 0.070 | 0.186 | 0.014 | 0.136 | 0.242 | 3 | 17.9% | +0.081 | 0.305 |
| 58909–58934[c] | −0.290 | 0.009 | 0.071 | 0.144 | 0.032 | 0.231 | 0.283 | 3 | 30.8% | +0.213 | 0.369 |

The stochastic contribution to $u_\text{a/lab}$ is now determined according to the model presented in section 2.6, and a minor adjustment has been applied to $u_\text{b}$ for the re-evaluated servo error uncertainty (section 2.1). From March 2016 to March 2020, a total of 13 NICT-Sr1 evaluations in conjunction with reports of eight Cs-fountain PFSs provide 63 datapoints $y(\text{Sr/SI})$ used to calculate the [87]Sr absolute frequency.

*2.11 Uncertainty contributions*

For each of these datapoints, we identify 10 uncertainty contributions:

$u_\text{a}^\text{PFS}$ and $u_\text{a}^\text{Sr}$ are the statistical uncertainties of the measurements $y(\text{HM}_\beta/\text{PFS})$ and $y(\text{HM}_\alpha/\text{Sr})$ resulting from the frequency standard itself. The corresponding errors are uncorrelated across separate measurements, but each measurement of NICT-Sr1 is compared to the results of multiple PFSs.

$u_\text{b}^\text{PFS}$ and $u_\text{b}^\text{Sr}$ are the systematic uncertainties of the frequency standards, which we take to be fully correlated over all measurement of the standard for the sake of calculating the uncertainty. For Cs-fountains, $u_\text{b}^\text{PFS}$ largely represents the individual characterization of systematic effects such as cavity phase distribution, atomic density and the blackbody radiation environment. As also adopted elsewhere [10,11,16,23], we thus consider no correlation of $u_\text{b}^\text{PFS}$ across different PFSs.

$u_\text{l/Tai}^\text{PFS}$ and $u_\text{l/Tai}^\text{Sr}$ describe the uncertainties for the satellite link of the local reference oscillator to TAI according to equation (12). Any constant equipment delays do not affect the frequency comparisons, and thus the uncertainties are taken to be uncorrelated for different periods. However, PFSs operated by the same institute share the same link, and there we assign a correlation coefficient of $T_\text{ov}/\sqrt{T_\text{a} \cdot T_\text{b}}$ according to the length of the overlap $T_\text{ov}$ in relation to the geometric mean of the individual intervals $T_\text{a}$ and $T_\text{b}$ [34].

$u_\text{a/lab}^\text{Sr}$ and $u_\text{b/lab}^\text{Sr}$ characterize the additional uncertainty in transferring the measurements of the frequency standard to the flywheel oscillator measured against TAI, as discussed in section 2.8. For the PFS, we take the reported value $u_\text{l/lab}^\text{PFS}$ to predominantly represent the deadtime uncertainty of the extrapolation $y(\mathrm{HM}_\beta, T_{\beta 2}/\mathrm{HM}_\beta, T_{\beta 1})$ in equation (16) and to be statistically independent between measurements, as is $u_{\mathrm{a/lab}}^{\mathrm{Sr}}$ for NICT-Sr1. We consider $u_{\mathrm{b/lab}}^{\mathrm{Sr}}$ to be fully correlated across all measurements.

The final contribution $u_{\mathrm{ext}}$ describes the uncertainty of extrapolating from interval $T_{\alpha 2}$ to $T_{\beta 2}$ using EAL as flywheel with an instability of

$$\sigma_y^2(\tau) = 3 \times 10^{-30}(\tau/\mathrm{d})^{-1} + 12 \times 10^{-32} + 4 \times 10^{-34}(\tau/\mathrm{d}) \qquad (19)$$

up to MJD 57809, and

$$\sigma_y^2(\tau) = 2 \times 10^{-30}(\tau/\mathrm{d})^{-1} + 9 \times 10^{-32} + 4 \times 10^{-34}(\tau/\mathrm{d}) \qquad (20)$$

after that. These values are given monthly in the report "Fractional frequency of EAL from primary frequency standards" [18] and we calculate a PSD $S_y^{\mathrm{EAL}}$ for the combination of white frequeny noise, flicker frequency noise and random-walk frequency modulation according to table 2. This typically introduces an uncertainty of several $10^{-16}$ unless evaluation periods agree exactly. A contribution from the estimated drift of EAL is generally one order smaller. Where extrapolations to different PFS evaluation periods extend over the same time span, the instability of EAL will introduce the same errors. For each datapoint, intervals that are evaluated by either the PFS or NICT-Sr1 (but not both) are considered part of the EAL extrapolation, and we assign a correlation coefficient for pairs of datapoints in the same way as for $u_{\mathrm{l/Tai}}$ above.

*2.12 Covariance and weights*

Given the vector **y** constructed from the 63 individual values $y(\mathrm{Sr}/\mathrm{SI})$, the mean value weighted according to the column vector of normalized weights **w** is simply $\bar{y}(\mathrm{Sr}/\mathrm{SI}) = \mathbf{w}^{\mathrm{T}}\mathbf{y}$. The corresponding uncertainty is described by a variance

$$\sigma^2 = \mathbf{w}^{\mathrm{T}} \mathbf{C} \mathbf{w} \;, \qquad (21)$$

where the $63 \times 63$ covariance matrix **C** describes the error correlations. Here and in the following, superscripts of T and $-1$ mark transposal and inversal.

We construct **C** as the sum of individual matrices $\mathbf{C}_i$ for each of the ten uncertainty contributions discussed in the previous section. Each $\mathbf{C}_i$ is simply constructed by combining the reported uncertainties with a correlation matrix containing coefficients of 0 for uncorrelated contributions and 1 for correlated contributions. For example, the correlation matrix for the NICT-Sr1 systematic uncertainty contribution consists entirely of ones, which results in an uncertainty that does not decrease with the addition of more data. Fractional correlation coefficients are assigned where there is a partial overlap of measurement data [34].

For a known covariance matrix **C**, the optimization problem to find optimal weights $\mathbf{w}_{\mathrm{op}}$ is solved by the Gauss-Markov theorem:

$$\mathbf{w}_{\mathrm{op}}^{\mathrm{T}} = \sigma_{\mathrm{op}}^2 \, \mathbf{j}^{\mathrm{T}} \mathbf{C}^{-1} \;, \text{ with } \; \sigma_{\mathrm{op}}^2 = (\mathbf{j}^{\mathrm{T}} \mathbf{C}^{-1} \mathbf{j})^{-1} \qquad (22)$$

The design matrix **j** is the column vector of ones $\mathbf{j} = (1,1,\ldots,1)^{\mathrm{T}}$. $\mathbf{w}_{\mathrm{op}}$ is optimal under the condition that knowledge of **C** is complete. We find a slight complication in that the conservative assumption of strongly correlated errors leads to a very uneven distribution of weights among repeated measurements referencing the same PFS: If the systematic clock error is reproduced perfectly in every measurement, then incorporating multiple measurements in the evaluation has no intrinsic value. The algorithm then concentrates almost all weight on the datapoints with the lowest additional error contributions from e.g. link uncertainty and extrapolation.

We consider this undesirable as it increases the sensitivity of the mean value to undiagnosed frequency excursions during the over-weighted measurements, and because even the type b uncertainties of the frequency standards are expected to have time-varying components that would tend to average out, even if this is not reflected in the uncertainty budget. For this reason, we apply a formally sub-optimal, but more even distribution of weights. We temporarily assign a reduced correlation coefficient of 0.5 for the systematic uncertainty of the same PFS in separate measurements to calculate a hypothetical covariance matrix $\tilde{\mathbf{C}}$. This is used only to determine a set of weights $\mathbf{w}'$ according to equation. (22) and find $\bar{y}(\mathrm{Sr}\,/\,\mathrm{SI}) = \mathbf{w}'^\mathrm{T}\,\mathbf{y}$ along with the accompanying uncertainty $\sigma^2 = \mathbf{w}'^\mathrm{T}\,\mathbf{C}\,\mathbf{w}'$, where the original, conservative covariance matrix $\mathbf{C}$ is used. Although $\mathbf{w}'$ is not formally optimal for this matrix, the increase in uncertainty is minimal, as shown below. The distribution of weights across PFSs and evaluation periods is included in figure 4.

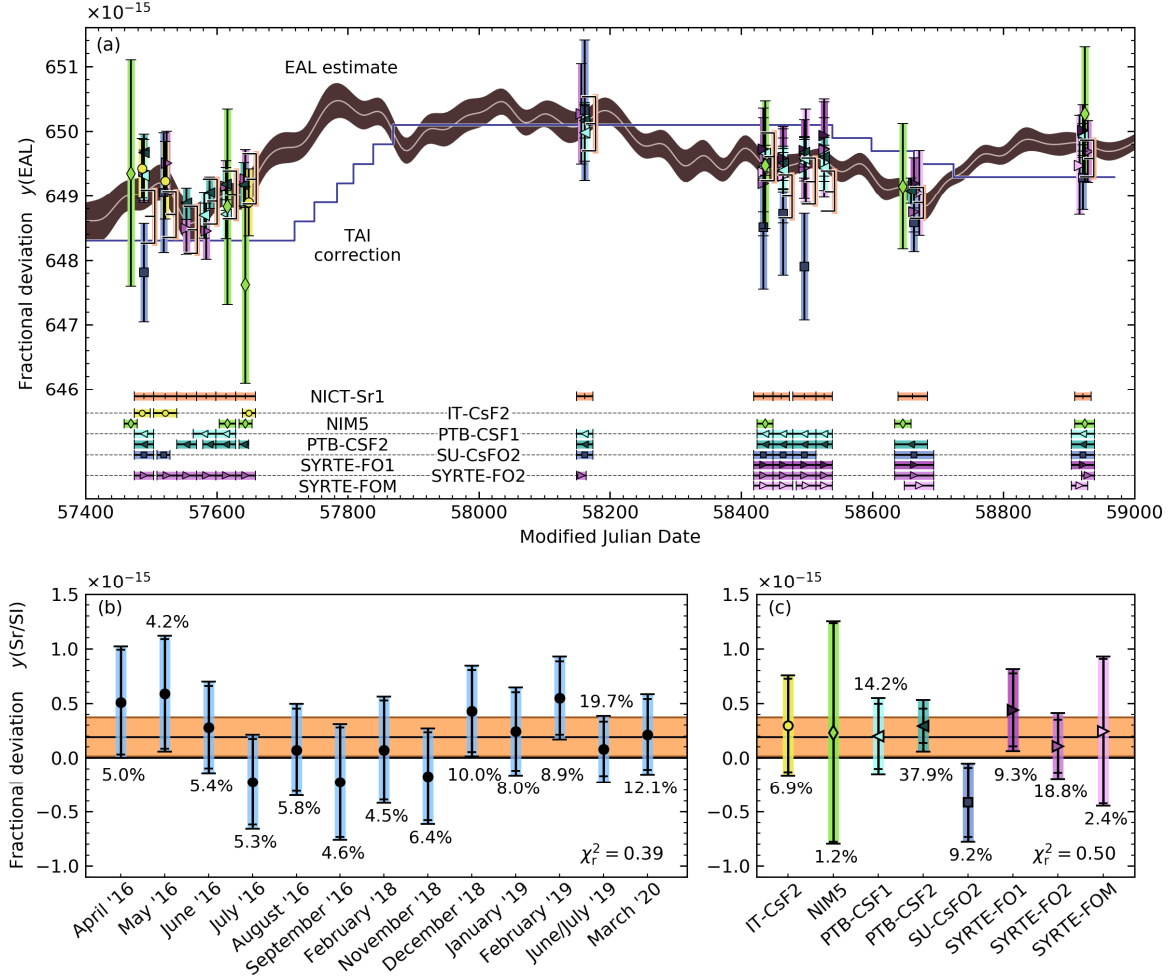

**Figure 4.** Overview of results. For 13 TAI evaluation periods of NICT-Sr1, all 63 overlapping PFS reports (coloured symbols) were evaluated to find the absolute frequency. **(a)** The reported deviations of the TAI scale interval $d$ were converted to a deviation $y(\mathrm{EAL})$ to remove the stepwise corrections applied to TAI (line). Error bars indicate total $1\sigma$ uncertainties of the evaluation. The results of NICT-Sr1 are shown by offset brackets for visibility over the PFS data points. The darkly shaded region indicates the $1\sigma$ uncertainty band for the BIPM estimate of $y(\mathrm{EAL})$, interpolated to illustrate the continuous behaviour. Horizontal bars at the bottom indicate the evaluation periods of the frequency standards relevant to this evaluation, spanning March 2016 to March 2020. **(b)** Averaged results for individual NICT-Sr1 evaluations in terms of the fractional deviation $y(\mathrm{Sr}/\mathrm{SI})$ from the nominal value $f_{\mathrm{SRS}}(\mathrm{Sr})$. Error bars indicate overall uncertainty, with additional smaller marks indicating the reduced variance $v_{\mathrm{rv}}$ expected relative to the weighted mean, as described in section 3. Percentages give the contributions to the overall mean displayed as the black line, with the orange-shaded region indicating a $1\sigma$ uncertainty of $1.8 \times 10^{-16}$. **(c)** Results grouped by referenced PFS and averaged over evaluation periods. Percentages represent relative contributions. Error bars indicate overall uncertainty of the average, with smaller marks indicating $v_{\mathrm{rv}}$ relative to the mean. Although the results for SU-CsFO2 show a larger than expected deviation, the reduced $\chi_{\mathrm{r}}^2$ is 0.50.

## 3. Results

Applying $\mathbf{w}'$ to the matrices $\mathbf{C}_i$ representing individual uncertainty contributions yields the uncertainty budget in table 4. The systematic uncertainties of the eight contributing PFSs give a combined $\bar{u}_b^{\text{PFS}} = 1.03 \times 10^{-16}$, similar to the values obtained in recent evaluations that trace the frequency of optical standards to TAI [10,11,16].

**Table 4.** Uncertainty budget of frequency evaluation.

| Contribution | Symbol | NICT-Sr1 ($\times 10^{-16}$) | PFS ($\times 10^{-16}$) |
|---|---|---|---|
| Clock statistical unc. | $\bar{u}_a$ | 0.06 | 0.27 |
| Clock systematic unc. | $\bar{u}_b$ | 0.72 | 1.03 |
| Link to local reference | $\bar{u}_{a/\text{lab}}$ | 0.68 | 0.12 |
| systematic contribution | $\bar{u}_{b/\text{lab}}$ | 0.60 | n.a. |
| Satellite link to TAI | $\bar{u}_{\text{TAI}}$ | 0.65 | 0.34 |
| Clock evaluation totals | | 1.33 | 1.13 |
| EAL extrapolation | $\bar{u}_{\text{ext}}$ | 0.35 | |
| Overall uncertainty | $\bar{u}_{\text{Sr}}$ | 1.78 | |

The overall fractional uncertainty of the mean is $\bar{u}_{\text{Sr}} = 1.78 \times 10^{-16}$ for the more evenly distributed weights $\mathbf{w}'$. Applying the formally optimal weights $\mathbf{w}_{\text{op}}$ would only result in a reduction to $1.75 \times 10^{-16}$.

Over the full set of datapoints, we find a reduced $\chi_r^2 = 0.45$, although this does not account for the expected correlations of the results. As an alternative measure of overall statistical consistency, we consider a reduced variance in the form of vector $\mathbf{v}_{\text{rv}} = \text{diag}(\mathbf{C}) - \mathbf{C}\,\mathbf{w}'$. This describes individual variances $v_{\text{rv}\_i}$ corrected for covariance with the correctly weighted mean. Applying $\mathbf{w}'$ to the standardized residuals $\Delta y_i^2 / v_{\text{rv}\_i}$ then gives

$$X_{\text{rv}}^2 = \sum_i w_i' \frac{\Delta y_i^2}{v_{\text{rv}\_i}} = 0.42 \;, \tag{23}$$

which is likewise consistent with a purely statistical variation. As shown in figure 4, we can also separate the results into contributions from individual evaluation of NICT-Sr1 and for individual PFSs. Averaged over evaluations, we find $\chi_r^2 = 0.39$ ($X_{\text{rv}}^2 = 0.38$), while over PFSs $\chi_r^2 = 0.50$ ($X_{\text{rv}} = 0.56$). Although we observe the contribution of SU-CsFO2 to deviate by $1.7\sigma$ from the mean, such a deviation is expected with 55% probability over the eight PFS contributions. Overall, the uncertainties listed in the Circular T appear to be rather conservatively estimated.

We then find a weighted mean $\bar{y}(\text{Sr/SI}) = 1.92(1.78) \times 10^{-16}$. This fractional deviation from the nominal value $f_{\text{SRS}}(\text{Sr}) = 429\,228\,004\,229\,873.0$ Hz in equation (17) represents an absolute [87]Sr frequency of $f(\text{Sr/SI}) = 429\,228\,004\,229\,873.082(76)$ Hz obtained by comparisons of NICT-Sr1 to the PFSs contributing to TAI calibration.

## 4. Discussion

This new evaluation is consistent with our earlier absolute frequency measurements [15,23,26] and improves on the most recent result with a reduction in overall uncertainty by a factor of 2.4. Since there have been only limited changes to the evaluation of systematic frequency shifts in NICT-Sr1, and parts of the evaluated period overlap, the new result should be considered a replacement of earlier data rather than an independent contribution.

Two other institutes have recently reported updates to their determination of the Sr frequency (table 5), by comparison to TAI [35], and against local PFSs [16]. Some of the most stringent tests of optical clocks now take the form of loop closures [10,36]. The Yb/Sr frequency ratio has recently been redetermined [37] as $\mathcal{R} = 1.207\,507\,039\,343\,337\,848\,2(82)$, with $7 \times 10^{-18}$ fractional uncertainty. Even though this differs by $1.8\sigma$ from previous results [38], the fractional difference is only $8 \times 10^{-17}$, significantly below the uncertainties of the absolute frequency measurements. This allows us to additionally consider two new results for the absolute frequency of the $^{171}$Yb clock transition [10,11]. The comparison values are included in table 5 together with corresponding $\tilde{f}_{Sr} = f_{Yb}/\mathcal{R}$, where $\mathcal{R}$ now contributes negligible uncertainty. Both Yb results also reference the ensemble of PFSs contributing to TAI calibration, and for the period of their determination state corresponding uncertainties $\bar{u}_b^{PFS}$ of $1.3 \times 10^{-16}$ and $1.2 \times 10^{-16}$. Taking the remaining contributions to be statistically independent, we find a mean value $\bar{f}_{Yb} = 518\,295\,836\,590\,863.67(10)$ Hz. With $\mathcal{R}$ and our value for $f_{Sr}$, this yields a misclosure

$$\delta = \frac{\bar{f}_{Yb}/f_{Sr}}{\mathcal{R}} - 1 = -1.4(2.6) \times 10^{-16} \quad , \tag{24}$$

well within statistical expectations. It is reasonable to expect that systematic frequency errors in the near-identical ensemble of referenced PFSs are largely identical for all three evaluations. A complete cancellation would reduce the uncertainty to $2.0 \times 10^{-16}$, which remains consistent with the observed misclosure.

**Table 5.** Evaluations of the $^{87}$Sr absolute clock transition frequency. Brackets indicate values calculated from recent absolute frequency measurements of the $^{171}$Yb clock transition.

| Source | | Ref. | $^{87}$Sr absolute frequency | Additional information |
|---|---|---|---|---|
| NICT | (JP) | (this) | 429 228 004 229 873.08(08) Hz | $1.8 \times 10^{-16}$ fractional uncertainty |
| PTB | (DE) | [16] | 429 228 004 229 873.00(07) Hz | $1.5 \times 10^{-16}$ fractional uncertainty |
| NPL | (UK) | [35] | 429 228 004 229 873.1(5) Hz | $1.2 \times 10^{-15}$ fractional uncertainty |
| CIPM 2017 | | [7] | 429 228 004 229 873.00(17) Hz | Secondary representation of the second |
| JILA | (US) | [37] | [429 228 004 229 873.04(09) Hz] | $\mathcal{R} = 1.207\,507\,039\,343\,337\,848\,2(82)$ |
| Yb: NIST | (US) | [2] | (see above) | $f_{Yb} = 518\,295\,836\,590\,863.71(11)$ Hz |
| Yb: INRIM | (IT) | [11] | [429 228 004 229 872.99(11) Hz] | $f_{Yb} = 518\,295\,836\,590\,863.61(13)$ Hz |

## 5. Conclusion

Figure 5 illustrates the excellent agreement of the various results, as well as the progression of uncertainties of the CIPM recommended frequencies for neutral strontium and ytterbium as Secondary Representations of the Second. Results as presented here, finding agreement between international measurements both in absolute frequency measurement and through the growing matrix of optical-to-optical comparisons, now support a further uncertainty reduction. If the CIPM were to adopt a new recommendation for the $^{87}$Sr clock transition frequency that reduces the uncertainty $u_{SRS}$ from its present value of $4 \times 10^{-16}$ to $1.8 \times 10^{-16}$, as we find in our evaluation, it would permit optical clocks like NICT-Sr1 (with average $u_b = 7.3 \times 10^{-17}$) to contribute to the steering of TAI with an effective systematic uncertainty $u_{b/eff} = \sqrt{u_b^2 + u_{SRS}^2} < 2 \times 10^{-16}$. This outperforms all PFSs considered in this evaluation apart from PTB-CSF2 and IT-CsF2 (both of which report $u_b = 1.7 \times 10^{-16}$), establishing optical clocks as first tier members of the TAI steering ensemble.

Directly tracing a frequency chain to individual PFSs not only helps minimize the uncertainty from mismatched evaluation periods, but also allows us to clearly state the fractional weights assigned to each standard, as included in Figure. 4. This will facilitate calculations that need to account for correlations between results [34]. As optical clocks gain increased weight in the steering of TAI, additional care is also required to make sure absolute frequency determinations are correctly traced to the SI second and the PFSs that implement it according to definition.

Our results were obtained using the same measurements, frequency links and communication channels used in the steering of the international time scale. Where local clock comparisons rely only on the local infrastructure, our results demonstrate accuracy and stability for the complete measurement chain.

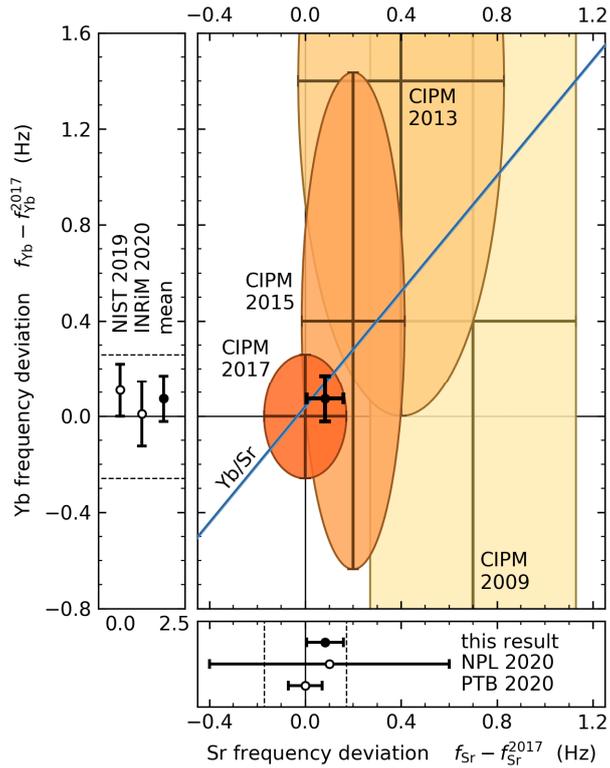

**Figure 5.** Comparison of results. The main figure shows the progression of the CIPM recommended frequencies for [87]Sr (horizontal) and [171]Yb (vertical) as coloured ellipses. All values are plotted relative to the 2017 CIPM recommendation. Measurements of the absolute frequencies [10,11, 16,35] reported later than this are shown by open circles and error bars to the left and below. The result reported here (solid point below) is in good agreement with the 2017 CIPM recommendation (dashed lines indicate 1σ confidence interval) and the recent measurements. The diagonal blue line in the main figure represents the confidence interval of a precise optical-to-optical measurement of the Yb/Sr frequency ratio [37]. Using this ratio to relate the mean of the Yb results (vertical error bar, see text) to our measurement of the Sr frequency (horizontal error bar), we also find agreement within statistical expectations.

## Acknowledgements

We are grateful to the many colleagues in time and frequency metrology who maintain and operate frequency standards, frequency links and the international timescales. We also thank BIPM for collecting all relevant information for public access. This work is enabled by their tireless efforts. Y. Li, H. Ishijima and S. Ito have been of great help in the construction and operation of NICT-Sr1 and we are fortunate to have their support. We also thank Piotr Morzyński for assistance in a long campaign and Marco Pizzocaro for inspiring discussions of deadtime-heavy measurements.